\documentclass{article}

\usepackage{PRIMEarxiv}

\usepackage[utf8]{inputenc} 
\usepackage[T1]{fontenc}    
\usepackage{hyperref}       
\usepackage{url}            
\usepackage{booktabs}       
\usepackage{amsfonts}       
\usepackage{nicefrac}       
\usepackage{microtype}      
\usepackage{lipsum}
\usepackage[dvipsnames,table,xcdraw]{xcolor}
\usepackage{soul}
\usepackage{amsmath,amssymb,amsfonts}%
\usepackage{fancyhdr}       
\usepackage{graphicx}       
\usepackage{ifluatex}
\usepackage{tablefootnote}
\usepackage{subcaption}
\usepackage{mathtools}
\usepackage{verbatim}
\usepackage{multirow}%

\usepackage{ulem}
\usepackage{setspace}
\ifluatex
  \usepackage{pdftexcmds}
  \makeatletter
  \let\pdfstrcmp\pdf@strcmp
  \let\pdffilemoddate\pdf@filemoddate
  \makeatother
\fi

\usepackage[notransparent]{svg}

\graphicspath{{media/}}     

\title{Comparing Self-Supervised Learning Techniques for Wearable Human Activity Recognition
}

\author{Sannara EK \\ sannara.ek@univ-grenoble-alpes.fr \\ University Grenoble Alpes \\ LIG F-38000 \\ Grenoble, France 
\And
\textbf{Riccardo PRESOTTO} \\ riccardo.presotto@unimi.it \\ University of Milan \\ EveryWare Lab \\ Milan, Italy 
\And
\textbf{Gabriele CIVITARESE} \\ gabriele.civitarese@unimi.it \\ University of Milan \\ EveryWare Lab \\ Milan, Italy 
\AND
\textbf{François PORTET} \\ francois.portet@imag.fr \\ University Grenoble Alpes \\ LIG F-38000 \\ Grenoble, France 
\And
\textbf{Philippe LALANDA} \\ philippe.lalanda@imag.fr\\ University Grenoble Alpes \\ LIG F-38000 \\ Grenoble, France 
\And
\textbf{Claudio BETTINI} \\ claudio.bettini@unimi.it \\ University of Milan \\ EveryWare Lab \\ Milan, Italy 
}

\begin{document}
\maketitle

\begin{abstract}
Human Activity Recognition (HAR) based on the sensors of mobile/wearable devices aims to detect the physical activities performed by humans in their daily lives. Although supervised learning methods are the most effective in this task, their effectiveness is constrained to using a large amount of labeled data during training. While collecting raw unlabeled data can be relatively easy, annotating data is challenging due to costs, intrusiveness, and time constraints.

To address these challenges, this paper explores alternative approaches for accurate HAR using a limited amount of labeled data. In particular, we have adapted recent Self-Supervised Learning (SSL) algorithms to the HAR domain and compared their effectiveness. We investigate three state-of-the-art SSL techniques of different families: contrastive, generative, and predictive. Additionally, we evaluate the impact of the underlying neural network on the recognition rate by comparing state-of-the-art CNN and transformer architectures.

Our results show that a Masked Auto Encoder (MAE) approach significantly outperforms other SSL approaches, including SimCLR, commonly considered one of the best-performing SSL methods in the HAR domain.

The code and the pre-trained SSL models are publicly available for further research and development.
\end{abstract}


\section{Introduction}
\label{intro}

Human Activity Recognition (HAR) aims to recognize human actions using sensor data automatically~\cite{chen2012sensor}. For simple physical activities, inertial sensors embedded into wearable devices, such as accelerometers and gyroscopes, are often preferred. The most effective approaches, in this case, rely on deep learning, where high recognition rates are achieved through the use of supervised learning techniques~\cite{wang2019deep}. This method, however, requires a significant amount of labeled data for model training and poses a real challenge due to the costs, intrusiveness, and time-consuming nature of data collection and annotation. Furthermore, the learned models obtained using data collected from certain users do not generalize well to other users~\cite{qin2019cross}. This effect is due to the idiosyncratic nature of HAR, as there are substantial variations in data distributions caused by factors such as differences in the way users perform activities, device placements, or sensing devices themselves.

We propose that these problems can be mitigated by promoting the use of robust pre-trained models \cite{han2021pre}, a foundational aspect of transfer learning where knowledge gained while solving one problem is applied to a different but related problem. While conventional classification problems require hard labels, we may first train the model on another given pretext task that does not necessarily require human-annotated labels and takes the same sensor inputs. Afterward, a portion or a layer of the network responsible for capturing relevant features from the input data, commonly referred to as the feature extractor, are combined with new or different model layers that are tasked to solve the targeted domain. This approach proves particularly advantageous when dealing with limited datasets or computational resources. However, the quality of the feature extractor heavily depends on the nature of the pretext task and its ability to leverage unlabeled data effectively.

For this purpose, Self-Supervised Learning (SSL) methods~\cite{van2020survey} can prove highly effective. The underlying principle of SSL is to establish alternative learning goals without reliance on labels, allowing pre-training on an abundant amount of data. Recent studies have identified three categories of trends in defining such goals through pretext tasks. Contrastive techniques involve training the model to minimize or maximize similarities or differences between samples, often done using data augmentations. Masking techniques, on the other hand, provide the model with only a partial view of the input data, compelling it to reconstruct the complete input data view. Predictive techniques require the model to generate subsequent data when provided with only past or present views of the input data. 

Recently, many research groups have been investigating the application of SSL to HAR. In this field, the challenge is to define a pretext task that reliably captures the spatiotemporal characteristics of sensor data. A recent survey~\cite{haresamudram2022assessing} suggests that SimCLR~\cite{chen2020simple} is one of the better-performing SSL methods for HAR.
However, SimCLR relies on selecting contrastive samples through data augmentations, which is known to be challenging~\cite{kalantidis2020hard}. Moreover, more recent approaches proposed in the other domains, like the Masked AutoEncoder (MAE)~\cite{he2022masked} and data2vec~\cite{baevski2022data2vec}, represent promising directions beyond contrastive learning and, at the same time, allows higher recognition rates. To the best of our knowledge, these methods have never been fully adapted and evaluated in the HAR domain.

The quality of each SSL method primarily depends on the quality of the pretext tasks. The pretext task must capture useful information to ensure the learned representations are meaningful. The purpose of our work is to thus adapt and compare recent state-of-the-art SSL techniques derived from computer vision (i.e., MAE and data2vec) to HAR and establish a relevant evaluation framework. This adaptation proved to be far from trivial and necessitated substantial technical and conceptual efforts. In particular, since SSL techniques are usually closely tied to the network's architecture, we conducted the implementation with two widely employed architectures: CNNs and Transformers. Finally, to obtain a larger diversified pre-training dataset and to extensively evaluate the trained models, we employ the Leave-One-Dataset-Out (LODO) method~\cite{presotto2023combining}. The method combines several publicly available HAR datasets, where one dataset is designated as the target domain, while the remaining ones are used to create the SSL pre-trained models. The process involves several folds, where we interchange the left-out dataset to present different sets of training datasets at each turn. In all scenarios, the pre-trained models were evaluated considering different labeled data scarcity scenarios, ranging from very few to large numbers. The specific contributions of this article are thus the following:

\begin{itemize}
    \item Adaptation of SSL methods of three different categories from the computer vision domain, namely Contrastive Learning (SimCLR), Generative Learning (MAE), and Predictive Learning (data2vec), specifically for the wearable Human Activity Recognition (HAR) domain.
    
    \item Conducting the first comprehensive evaluation of these three SSL categories together in the wearable HAR domain, considering both data-rich and data-scarce environments.
    
    \item The study is fully reproducible since all research results, code, and pre-trained models have been made publicly available on a repository for easy access and further exploration\footnote{https://github.com/getalp/Self-Supervised-Learning-HAR}.
    
    \item The findings indicate that MAE emerges as the most robust technique for the domain, surpassing all other approaches, including supervised pre-training methods.

\end{itemize}

The paper is organized as follows. First, some background is provided about HAR and SSL techniques. Next, we detail the process of obtaining multiple pre-trained models using adapted SSL techniques with two different network architectures. Based on the LODO method, our evaluation pipeline is presented in section 4. Section 5 reports experiments performed and is followed by a discussion about the obtained results. Finally, the paper concludes based on our main findings and an outlook on future work.

\section{Related Work}
\label{related}

Dealing with the scarcity of labeled data in the domain of HAR is a significant challenge. Researchers have investigated diverse methodologies to tackle this issue, enabling the application of advanced approaches to reduce the reliance on labeled data significantly.

One of these methods is Transfer learning\cite{hernandez2020literature,pavliuk2023transfer}, which leverages pre-existing models trained on large labeled datasets to mitigate the amount of fine-tuning data required for a new target domain. However, transfer learning still requires a significant amount of labeled data from the source domain.

Semi-supervised learning is another technique to mitigate labeled data scarcity, which uses a small amount of labeled data while leveraging a much larger amount of unlabeled data~\cite{4911590}. For example, in the active learning approach, the model requires the user to provide a label to the most informative unlabeled samples (i.e., the ones where the model is uncertain). These samples are then used to improve the model~\cite{presotto2022semi}. Other well-known semi-supervised approaches are self-learning~\cite{tang2021selfhar}, label propagation~\cite{stikic2009multi} and co-training~\cite{guan2007activity}. However, Semi-supervised learning methods still require labeled data to initialize the recognition model. Moreover, methods like active learning require bothering the users, while methods like self-learning are prone to error propagation.


Another method for mitigating data scarcity is fabricating synthetic data with generative models to enhance the limited labeled data available~\cite{chan2021unified}. However, synthetic data are often unrealistic and do not help the model generalize.

Recently, Self-Supervised Learning (SSL) has become one of the most popular methods to mitigate labeled data scarcity~\cite{haresamudram2022assessing}. SSL is a specific type of transfer learning where a feature extractor leverages large amounts of unlabeled data from the source domain. Here, the challenge is finding the pretext task the model should use to learn a reliable representation from unlabeled data only. There have been several proposals of pretext tasks for SSL~\cite{mohamed2022self}. Selecting the most suitable one requires careful consideration of the characteristics and structures of the input data. The most common categories of SSL approaches are generative, contrastive, and predictive. Each category is distinguished by its unique pretext task.
%
%
%


In generative approaches, the pretext task is the reconstruction of a masked portion of the input by only observing the unmasked part ~\cite{Devlin2019BERTPO}. Contrastive approaches use pairs of similar and dissimilar examples as inputs, and the pretext task is to learn a distance function, keeping close the similar examples and pushing apart the dissimilar ones~\cite{chen2020simple}. Hence, the feature extractor aims to create similar representations for similar data and significantly different ones for dissimilar data. Lastly, the pretext task of predictive approaches involves the prediction of some properties of data samples. For instance, pre-trained Large Language Models like GPT are trained using a pretext task to predict the following word in the text~\cite{ericsson2022self}.

Although SSL has primarily been explored in Natural Language Processing and vision, it has recently been applied to sensor-based HAR by several research groups~\cite{haresamudram2022assessing}. The majority of these works have focused on contrastive approaches ~\cite{khaertdinov2021contrastive,wang2022sensor,jain2022collossl,saeed2020federated,arrotta2023selfact}. A significant drawback of these methods is that selecting contrastive examples is a challenging and resource-intensive task. The complexity is because negative examples should be sufficiently ``hard'' to learn reliable feature representations~\cite{kalantidis2020hard} effectively. Other studies proposed generative methods ~\cite{haresamudram2020masked,xu2021limu,zhang2022cross} and demonstrated competitive performances. To the best of our knowledge, only a few predictive approaches proposed in the literature have been proposed ~\cite{haresamudram2021contrastive,tang2021selfhar}. A recent survey has shown that SimCLR~\cite{chen2020simple}, a contrastive approach, is the most effective method for reliable feature extraction ~\cite{haresamudram2022assessing}. 
However, this survey does not take into account the latest advancements in the field of vision and NLP, such as MAE~\cite{he2022masked} and data2vec~\cite{baevski2022data2vec}, which are presented later in this article.

\section{Adapting Self-Supervised Learning Methods for HAR}
\label{sec:SSL}

In this paper, we carefully selected and adapted the most effective algorithms from each major SSL category, leveraging SimCLR~\cite{chen2020simple} for contrastive learning, MAE \cite{he2022masked} for generative learning, and Data2vec~\cite{baevski2022data2vec} for predictive learning. In this section, we provide a comprehensive summary of these methods, including their underlying principles, limitations, distinctions from one another, and the specific adaptations we made to align them with the wearable HAR domain. Additionally, we considered two state-of-the-art architectures: iSPLInception (iSPL)~\cite{9425494}, a convolution-based network that incorporates an inception module from the vision domain, and Human Activity Recognition Transformer (HART)~\cite{ek2023transformer}, a lightweight sensor-wise transformer-based architecture.

\subsection{SimCLR}

SimCLR \cite{chen2020simple} is a contrastive learning method that learns meaningful patterns from unlabelled data. It works by taking a single input and applying different transformations, creating what we call 'positive pairs'. At the same time, it takes other transformed samples from the same batch and treats them as 'negative pairs'. Each of these transformed samples is then encoded to create embeddings. These embeddings are passed through what we call 'projection heads' for the actual contrastive learning process.

The goal of SimCLR's loss function is to make sure that the representations of the positive pairs (the ones that came from the same original input) are as similar as possible. In contrast, the representations of the negative pairs (the ones that came from different original inputs) are as different as possible from the positive pairs. This way, the model learns to distinguish between different kinds of inputs.

\begin{figure}[h]
     \centering
     \includegraphics[width=0.90\linewidth]{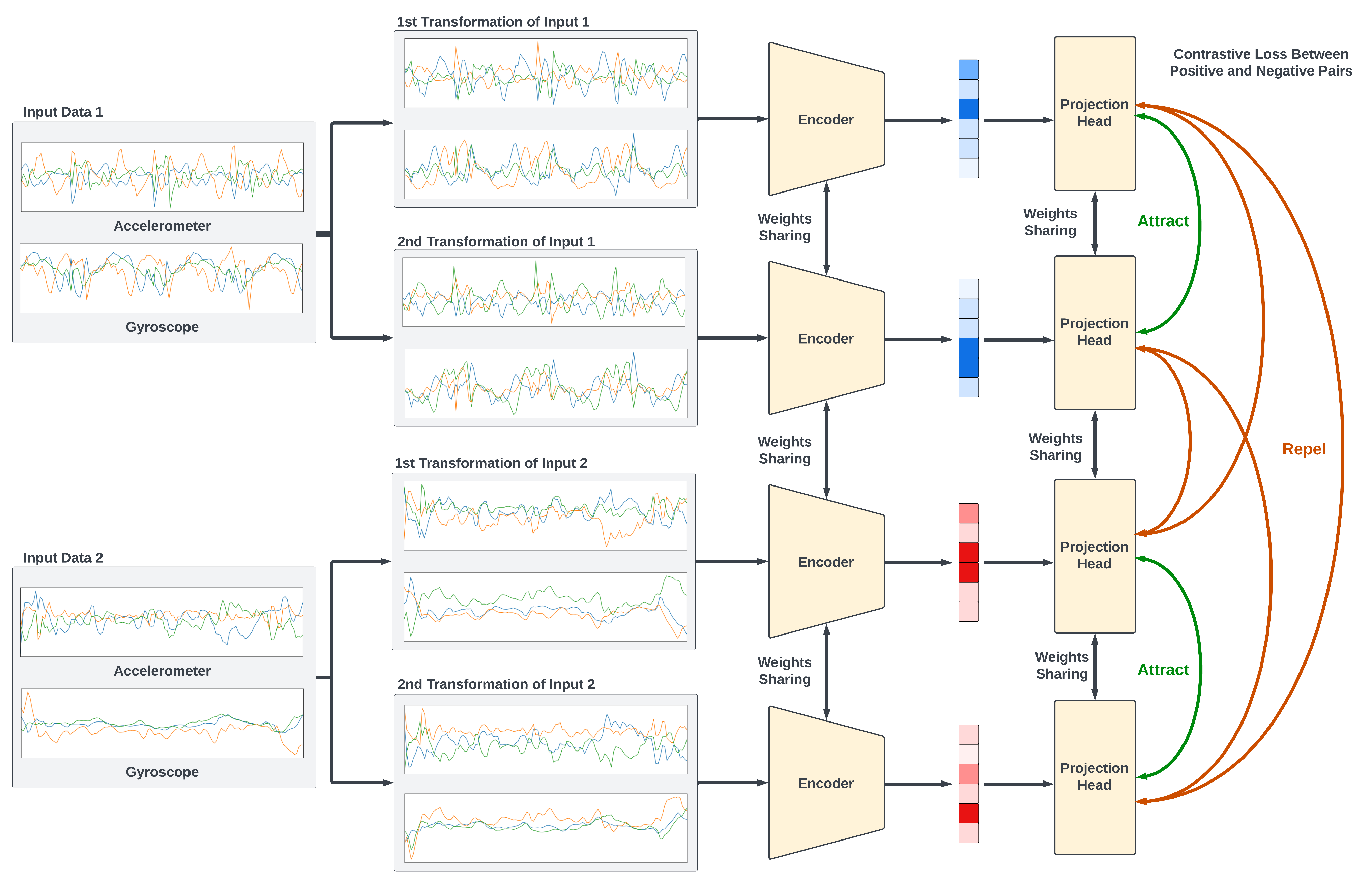}
    \caption{Overview of SimCLR for HAR}
   \label{fig:simCLR_METHOD}
\end{figure}

A previous study~\cite{tang2020exploring} already proposed a method to adapt SimCLR to the HAR domain, finding that random 3D rotation was an effective transformation for accelerometer data. The implementation of this approach is public, and we slightly extended it accommodate to multiple sensors. In our adaptation, both accelerometer and gyroscope data are separately subjected to 3D rotation and fed into the neural network for contrastive learning. Figure~\ref{fig:simCLR_METHOD} illustrates this process, depicting the augmentation process and its utilization as input data. Moreover, we adapted the SimCLR to use HART and iSPL networks as the encoders. For the projection heads, we used the one proposed in the original implementation~\cite{tang2020exploring}, which are 3 stacked dense layers of sizes 256, 128, and 50.

Despite its success, the SimCLR approach has certain limitations and methodological constraints. Firstly, data transformations are domain-specific. While operations like image rotation or flipping are intuitive in computer vision, they are less straightforward in other domains like HAR. Discovering appropriate transformations becomes a genuine challenge and is crucial for the method's success. Another significant limitation of SimCLR is that un-augmented data is never used for training. This issue is specifically detrimental during the downstream task, where the model underperforms on actual data due to significant differences in the transformed and original input data, regardless of having a converged contrastive loss.

\subsection{MAE}

\begin{figure}[h]
    \centering
    \includegraphics[width=\linewidth]{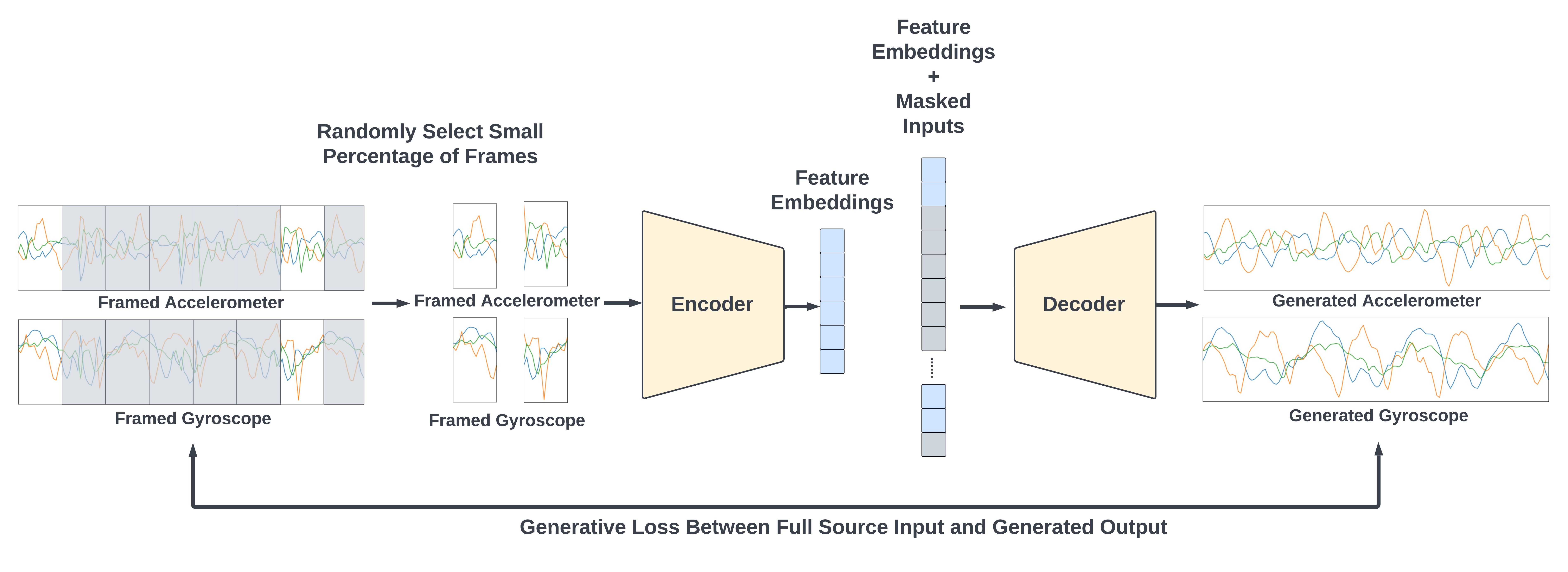}
    \caption{Overview of Masked Autoencoder (MAE) for HAR.}
    \label{fig:mae_METHOD}
\end{figure}

Masked Auto Encoders (MAEs) \cite{he2022masked}, as illustrated in Figure~\ref{fig:mae_METHOD}, is a generative/masking method designed to understand the underlying structure of unlabeled data by reconstructing it based on partial views. A linear projector first transforms the input data into a fixed-size representation. Most of the projected inputs are randomly selected and masked during this process. This masking is done using learnable mask tokens, which are vectors of the same length as the transformed input. These mask tokens are also trained through back-propagation. Only a small subset of the remaining frames, which are not masked, is used as input for the encoder. This lightweight mechanism allows for efficient processing. The output of the encoder is then combined with the masked projections and fed into a lightweight decoder.

MAE aims to minimize the difference between the original inputs and the reconstructed outputs. By reconstructing the original input with only a partial view, MAE effectively learns to capture and encode crucial information. However, one potential limitation of MAE is that it requires the model to reconstruct data in the original space, not in the latent space. This detail means the model will be required to reconstruct noisy low-level details.


We adapted MAE for HAR, building upon the original implementation \cite{he2022masked}, which was primarily designed for computer vision tasks. Our approach involved modifying the iSPL to process inputs as a series of non-overlapping, independent frames, similar to how transformers function. For the integration to HART, we introduced sensor-specific masking tokens to accommodate the sensor-wise feature of HART. We endeavored to stay true to the original implementation in all other aspects. To the best of our knowledge, this is the first integration of MAE into the HAR domain that is closely aligned with the original implementation.

\subsection{Data2vec}

\begin{figure}[h]
    \centering
    \includegraphics[width=0.9\linewidth]{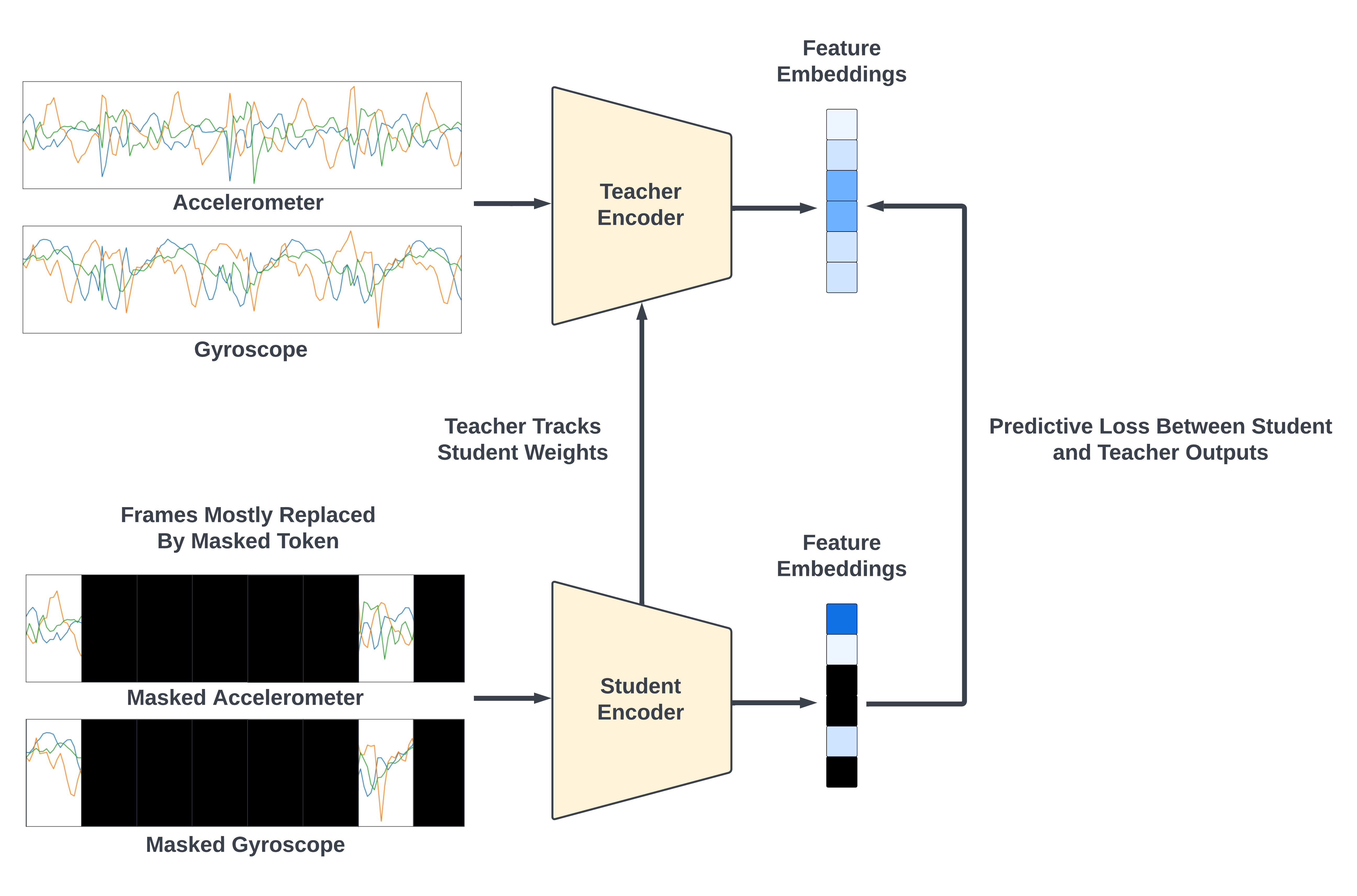}
    \caption{Overview of Data2vec for HAR  }
    \label{fig:data2vec_METHOD}

\end{figure}

Data2vec~\cite{baevski2022data2vec}, as depicted in Figure~\ref{fig:data2vec_METHOD}, aims to learn robust feature representations by predicting the latent representation of an input data sample from a partial view. The method is a unified framework designed for multiple domains, including vision, speech, and Natural Language Processing (NLP). Data2Vec is composed of a student and a teacher network. In this framework, the student model receives an augmented view of the input data, with most frames masked. In contrast, the teacher model operates on all input data, and its outputs serve as the training targets for the student.

The output of the teacher's encoder is computed as the average of the model's top layers, aiming to reduce the risk of producing outputs too similar to those generated by the student's model. In this student-teacher architecture, the student model is trained to predict only the teacher's contextualized representations for the masked time steps. Only the student model's weights are updated during training, while the teacher model is updated by computing the moving average of the student model's weights. A smooth L1 loss is used to minimize the distance between the embedding generated by the teacher and student encoders.


In a manner similar to MAE, each input data sample in data2vec must first be framed and then transformed into a fixed-size vector in a latent space using a linear projector. A majority of these frames are selected for masking. Unlike MAE, data2vec replaces the entirety of the masked frames with learnable mask tokens. These tokens and the unmasked projections are fed into the student encoder.

A significant limitation of data2vec is the risk of representation collapse, where the teacher's training target becomes trivial for the student to predict. This issue can hinder the model's learning process and is more common in teacher-student configurations like data2vec since the teacher's weights are a moving average of the students. Mitigating this effect necessitates more precise hyperparameter tuning to counteract this drawback.

As far as current knowledge extends, Data2Vec has yet to be applied to the HAR domain. Since there was no public implementation of data2vec with TensorFlow (i.e., the framework we used to implement the other methods), we implemented it from scratch, directly adapting it to the sensor-based HAR domain. Similarly to what we did for MAE, we modified iSPL and MAE to work with sensor data. Moreover, iSPL was adapted to process the inputs as non-overlapping frames.

\newpage
\section{Evaluation Strategy \& Experimental Setup}
\label{sec:eval_dataset}
In the wearable HAR domain, a significant challenge arises from the absence of large and diverse datasets that can serve as a suitable source for pre-training a generalized SSL model. Despite recent advancements~\cite{chan2021capture,doherty2017large}, there have yet to be publicly available Gold-standard datasets that are expansive and inclusive of diverse devices, positions, users, and sensors. Moreover, it is crucial to employ a realistic evaluation strategy that involves distinct source and target domains and effectively addresses the issue of data scarcity and the out-of-distribution problem. This section provides an overview of the datasets used and the evaluation strategies employed to assess the performance of the pre-trained models obtained through various SSL techniques and architectures.

\subsection{Leave-One-Dataset-Out}


\begin{figure}[h]
    \centering
    \includegraphics[width=\linewidth]{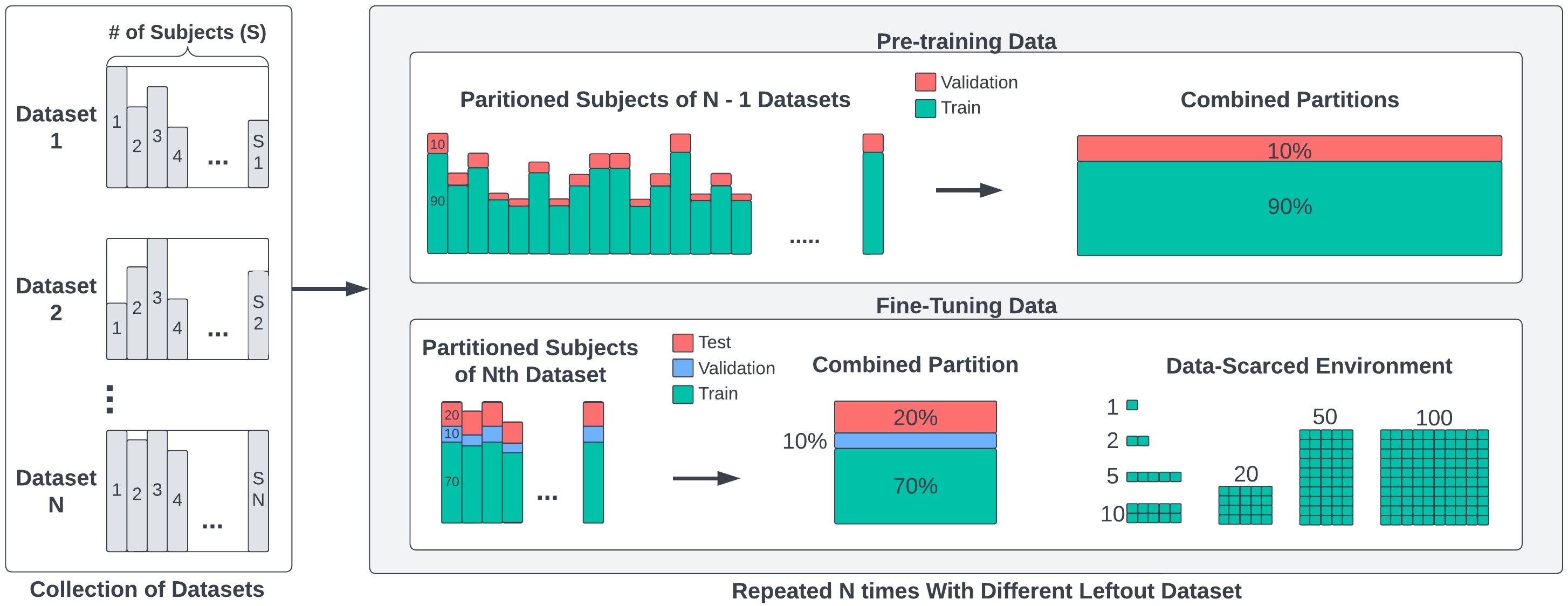}
    \caption{Overview of the Leave-One-Dataset-Out evaluation methodology}
    \label{fig:extended_LODO}
\end{figure}

To deal with limited dataset availability and ensure broad coverage on various properties and distribution, we have extended the Leave-One-Dataset-Out (LODO) evaluation strategy~\cite{presotto2023combining}. This strategy, as illustrated in figure \ref{fig:extended_LODO}, combines multiple public datasets, with one set aside as the target domain for each process iteration. The combined datasets are used to pre-train models using SSL techniques. The datasets are divided subject-wise, with 90\% of the data used for training and 10\% for validation. The training data from all subjects across the datasets are merged to form the source dataset, which is used to learn the pre-trained model.

The set-aside dataset, known as the target dataset, is also divided subject-wise. 20\% of this data is reserved for testing, 10\% for validation, and the remaining 70\% is used to create various training partitions. These partitions simulate different settings of labeled data scarcity for the fine-tuning process.

Unlike previous work where data scarcity was simulated using small percentages of the left-out dataset, this paper presents a more challenging learning environment. It employs a fixed small number of samples per class, specifically considering 1, 2, 5, 10, 20, 50, and 100 samples per class in the low data-resourced setting. We refer to each unique available sample count per class for fine-tuning as a scenario. Each scenario combines samples from the scenarios with fewer samples per class, adding additional new samples. In addition, a learning instance where the entirety of the training dataset is used for comparison with conventional approaches.

We fine-tune the target environment in two scenarios: one with the feature extractor frozen (encoder weight not updated during fine-tuning) and another with the extractor unfrozen (encoder weight updated during fine-tuning), following the approach used in previous work~\cite{presotto2023combining}. In both scenarios, we compare SimCLR, MAE, Data2vec, and conventional supervised learning as a baseline for the pre-training technique. Evaluating a model that has been fine-tuned with its features frozen allows for a better assessment of the capabilities of each pre-training technique studied. However, when the feature extractor is unfrozen, the benefits of pre-training are less comparable, as the pre-trained weights act as an `initializer' while the weights of the new model undergo significant changes. Nonetheless, the model fine-tuned with its feature extractor unfrozen tends to achieve better results, albeit with some additional efficiency cost.

\subsection{Datasets}

In the following, we describe the six datasets (HHAR, MobiAct, MotionSense, Realworld, UCI, and PAMAP2) employed in this study. It is essential to highlight that our focus was explicitly on datasets containing data from accelerometer and gyroscope sensors. We gathered seven datasets, each possessing unique traits that distinguish them from one another. The Heterogeneity Human Activity Recognition (HHAR) dataset~\cite{10.1145/2809695.2809718} showcases a diverse range of smartphone and smartwatch models, adding to our variability. MobiAct~\cite{vavoulas2016mobiact} includes 61 subjects, making it the dataset with the largest subject count in our collection. MotionSense~\cite{Malekzadeh:2018:PSD:3195258.3195260} exclusively comprises data collected from Apple iPhones, setting it apart from the rest. RealWorld~\cite{realword} brings in further diversity by incorporating seven on-body positions. The UCI Human Activity Recognition dataset~\cite{Anguita2013APD} has long served as a benchmark within the community for HAR, owing to its extensive usage. Finally, PAMAP2~\cite{reiss2012introducing} poses the most significant challenge among the datasets we studied, including 12 distinct physical activities.

All datasets were reduced to a sampling frequency of $50$ Hz to match the recommended frequency for HAR on smartphones~\cite{overviewHar}. After downsampling, each dataset was individually standardized using sensor-wise z-normalization to center the data. To avoid under-representing datasets with fewer samples when combined with larger datasets, we standardized them separately rather than together. This approach prevents the influence of domain shifts from different positions~\cite{ek2023transformer}. Specifically, we focused on the `waist' position since it is the most common and present in all the datasets used in this study.

Next, the data were segmented into windows which contain $128$ samples (equivalent to $2.56$ seconds) with a $50$\% overlap for the accelerometer and gyroscope readings across all six channels, as past studies have done~\cite{ignatov2018real}. When merging the datasets, we only considered the physical activities of each dataset, aligned them, and performed a union of all the filtered-out labels. This resulted in a dataset comprising $147$ subjects, $10$ unique activities (such as Downstairs, Upstairs, Running, Sitting, Standing, Walking, Lying, Cycling, Nordic Walking, and Jumping), with a total of $213,289$ data samples (approximately $151$ hours of usable data). The data was acquired using $12$ different devices. Despite considering only one body position, the diversity of subjects and devices in the combined data presents a challenging learning problem due to its heterogeneity.

\subsection{Experimental Setup}

We conducted experiments on a high-performance computing cluster with Intel Cascade Lake 6248 processors, 192GB of memory, and Nvidia Tesla V100 16GB GPUs. The development of all models was carried out using TensorFlow. For each experiment, the pre-training was done with $300$ epochs and a batch size of $128$. We implemented early stopping, where after 15 epochs of no further decrease in loss, we stopped the training and took the model with the lowest loss on the pre-training validation set. During the fine-tuning phase, we performed $100$ epochs with a batch size of $64$ on the left-out datasets. For the downstream environment, we only take the feature extractor of the pre-trained model and attach a dense layer of shape 1024, followed by the classification heads. The test set was evaluated using the fine-tuned model that attained the highest accuracy on the validation set. We employed an Adam optimizer with a learning rate of $0.0003$ for all experiments. The performance of the classification task was assessed using the macro F1 score metric.

Finally, in addition to the adaptations performed in this work, we conducted an extensive grid search involving three hyper-parameters for each of the presented SSL methods using two architectures on the HHAR dataset. We detail here that all encoders utilize default parameters from the two architectures across all SSL techniques to ensure fairness in the comparison. For the remaining hyper-parameters, we select those that yielded the lowest loss in each category, and our findings are presented in table \ref{table:gridsearch}.

\begin{table}[h]
\centering
\caption{Hyper-parameters chosen for the study obtained from a grid search}
\begin{tabular}{lcccccc}
Architecture                         & \multicolumn{3}{c}{iSPL}                       & \multicolumn{3}{c}{HART} \\ \hline
\multicolumn{1}{l|}{Method}          & SimCLR   & MAE & \multicolumn{1}{c|}{Data2vec} & SimCLR  & MAE & Data2vec \\
\multicolumn{1}{l|}{Batch Size}      & 128      & \_  & \multicolumn{1}{c|}{\_}       & 128     & \_  & \_       \\
\multicolumn{1}{l|}{Mask Ratio}      & \_       & 0.6 & \multicolumn{1}{c|}{0.75}     & \_      & 0.6 & 0.5      \\
\multicolumn{1}{l|}{Tau}             & \_       & \_  & \multicolumn{1}{c|}{0.998}    & \_      & \_  & 0.9999   \\
\multicolumn{1}{l|}{Beta}            & \_       & \_  & \multicolumn{1}{c|}{0.5}      & \_      & \_  & 0.5      \\
\multicolumn{1}{l|}{Decoder Depth}   & \_       & 4   & \multicolumn{1}{c|}{\_}       & \_      & 6   & \_       \\
\multicolumn{1}{l|}{Decoder Width}   & \_       & \_  & \multicolumn{1}{c|}{\_}       & \_      & 252 & \_       \\
\multicolumn{1}{l|}{Decoder Filters} & \_       & 192 & \multicolumn{1}{c|}{\_}       & \_      & \_  & \_       \\
\multicolumn{1}{l|}{Temperature}     & 0.1      & \_  & \multicolumn{1}{c|}{\_}       & 0.1     & \_  & \_       \\
\multicolumn{1}{l|}{Transformation}  & Rotation & \_  & \multicolumn{1}{c|}{\_}       & Noise   & \_  & \_      
\end{tabular}
\label{table:gridsearch}

\end{table}

\section{Results}
\label{sec:results}

In this section, we describe the results obtained using the evaluation methodology presented in Section~\ref{sec:eval_dataset}. 
Initially, we considered an optimistic scenario where all labeled data from the target domain are available and can be used to fine-tune the pre-trained model. This experiment aims to demonstrate the potential of transfer learning between a model pre-trained with SSL and a target dataset. Then, we also considered several data scarcity scenarios by varying the number of labeled data (per class) used to fine-tune the pre-trained model. In this case, the objective is to assess how the quality of embeddings can reduce the number of labeled data required for fine-tuning and to what extent. All configurations of our experiments have undergone $5$ random runs to prevent biased results due to randomness traits on the training pipelines.

\subsection{Fine-tuning with full training data}

In the following, we present the results of our experiments in the optimistic scenario with full availability of training data in the target domain for fine-tuning the pre-trained model in the downstream task. Note that since our experiments include many permutations of hyper-parameters (4 pretraining methods, 2 architectures, and 7 datasets), we have averaged the results on all the folds of the LODO evaluation. Table~\ref{table:averagedResultsFrozen} shows the results of the SSL approaches presented in Section~\ref{sec:SSL}, by using a \textit{frozen} setting during fine-tuning. This table shows that MAE significantly outperforms the other SSL approaches, including the supervised baseline. Moreover, SimCLR is the worst-performing method. Note that the high standard deviation is due to the fact that we consider many datasets with very different characteristics. Hence, the recognition rate on each left-out dataset varies based on its complexity. We provide dataset-level
results in Sections~\ref{subsec:datasetlevelfrozen} and ~\ref{subsec:datasetlevelunfrozen}.

\begin{table}[h]
\centering

\caption{Average F1-scores (± std) with the model’s feature extractor \textbf{frozen}}
\begin{tabular}{cll}
Methods                         & \multicolumn{1}{c}{iSPL} & \multicolumn{1}{c}{HART} \\ \hline
\multicolumn{1}{c|}{Supervised} & 65.74 ± 16,56            & 74.95 ± 15.92            \\
\multicolumn{1}{c|}{SimCLR}     & 62.96 ± 18.71           & 50.20 ± 16.65            \\
\multicolumn{1}{c|}{MAE}        & \textbf{80.90 ± 10.88}  & \textbf{82.56 ± 11.38}   \\
\multicolumn{1}{c|}{Data2vec}   & 66.91 ± 12.65            & 73.84 ± 15.35 \\ 
\hline
\end{tabular}
\label{table:averagedResultsFrozen}
\end{table}

\begin{figure}[h]
    \centering
    \includegraphics[width=\linewidth]{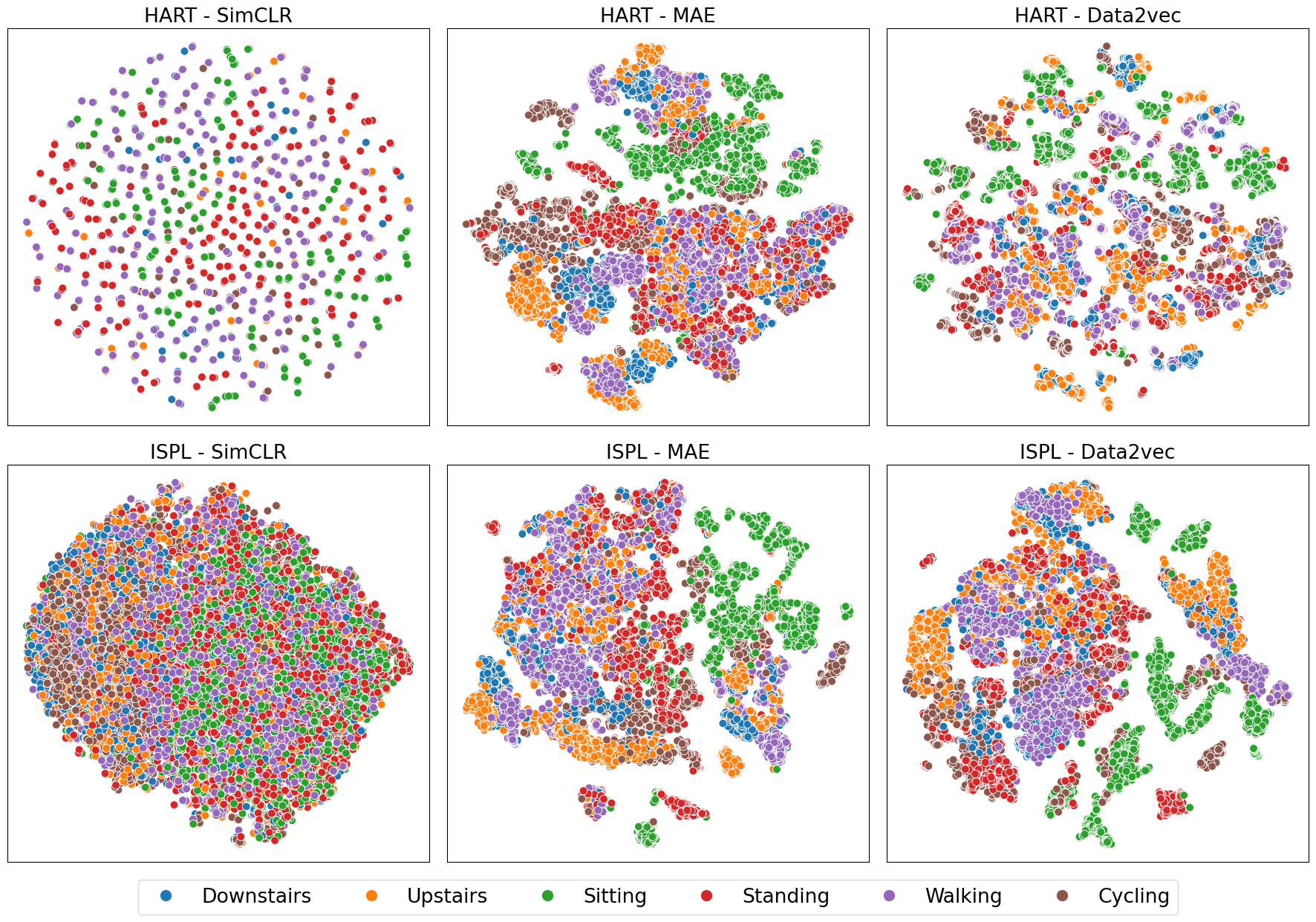}
    \caption{T-SNE projections of the studied SSL methods on the HHAR dataset}
    \label{fig:embedding}
\end{figure}

For a qualitative analysis in this aspect, we used T-SNE to project the embeddings of the three SSL techniques with the two different architectures on the HHAR dataset left out and present it in figure \ref{fig:embedding}. Findings are in line with the statements above; while MAE and Data2vec can form relatively separable clusters between activities, the embedding of SimCLR is either too conjoined or loosely separated.

The overall averaged result for the unfrozen setting is shown in table~\ref{table:averagedResultsUnfrozen}. We observe that there is no clear winner. Indeed, all SSL methods reach very close recognition rates independently from the underlying neural network. Moreover, the unfrozen setting reaches significantly higher recognition rates than the frozen one. The lack of difference between each SSL method is because the weights from pre-training are overridden when the whole network is trained using all the available training data.

\begin{table}[h]
\centering
\caption{Average F1-scores (± std) with the model's feature extractor \textbf{unfrozen}}
\begin{tabular}{cll}
Methods                         & \multicolumn{1}{c}{iSPL} & \multicolumn{1}{c}{HART} \\ \hline
\multicolumn{1}{c|}{Supervised} & 90.77 ± 9.18            & 89.50 ± 10.70            \\
\multicolumn{1}{c|}{SimCLR}     & \textbf{90.81 ± 8.57}          & 89.33 ± 9.72            \\
\multicolumn{1}{c|}{MAE}        & 88.14 ± 9.25             & \textbf{89.57 ± 8.92}   \\
\multicolumn{1}{c|}{Data2vec}   & 88.58 ± 8.62            &  89.31 ± 8.65   \\ 
\hline
\end{tabular}
\label{table:averagedResultsUnfrozen}
\end{table}

\subsubsection{Detailed Results in the Frozen case}
\label{subsec:datasetlevelfrozen}

\begin{table}[h]
\centering
\caption{Average F1-scores (± std) over 5 random runs with the model's feature extractor \textbf{frozen} for each dataset as a downstream task. Best results are reported in \textbf{bold}, while second-best are performances are \uline{underlined}}
\begin{tabular}{cccccc}
Archt.                                 & Datasets                         & Supervised                        & SimCLR                            & MAE                                        & Data2vec     \\ \hline
\multicolumn{1}{c|}{}                       & \multicolumn{1}{c|}{HHAR}        & \multicolumn{1}{c|}{39.05 ± 3.91} & \multicolumn{1}{c}{28.23 ± 1.18} & \multicolumn{1}{c}{\textbf{66.11 ± 1.05}} & \uline{47.43 ± 4.68} \\
\multicolumn{1}{c|}{}                       & \multicolumn{1}{c|}{MobiAct}     & \multicolumn{1}{c|}{61.46 ± 2.78} & \multicolumn{1}{c}{66.41 ± 1.48} & \multicolumn{1}{c}{\textbf{81.24 ± 0.59}} & \uline{70.90 ± 2.73} \\
\multicolumn{1}{c|}{}                       & \multicolumn{1}{c|}{MotionSense} & \multicolumn{1}{c|}{74.42 ± 1.87} & \multicolumn{1}{c}{\uline{83.75 ± 0.65}} & \multicolumn{1}{c}{\textbf{91.98 ± 0.38}} & 78.34 ± 2.36 \\
\multicolumn{1}{c|}{}                       & \multicolumn{1}{c|}{RealWorld}   & \multicolumn{1}{c|}{\uline{76.78 ± 2.25}} & \multicolumn{1}{c}{71.05 ± 1.98} & \multicolumn{1}{c}{\textbf{87.06 ± 0.44}} & 75.54 ± 4.10 \\
\multicolumn{1}{c|}{}                       & \multicolumn{1}{c|}{UCI}         & \multicolumn{1}{c|}{\uline{85.18 ± 1.88}} & \multicolumn{1}{c}{68.19 ± 1.80} & \multicolumn{1}{c}{\textbf{92.17 ± 0.83}} & 76.12 ± 4.01 \\
\multicolumn{1}{c|}{\multirow{-6}{*}{iSPL}} & \multicolumn{1}{c|}{PAMAP2}       & \multicolumn{1}{c|}{57.59 ± 0.78} & \multicolumn{1}{c}{\uline{60.16 ± 0.38}} & \multicolumn{1}{c}{\textbf{66.82 ± 1.48}} & 53.12 ± 4.07 \\
\multicolumn{6}{l}{\cellcolor[HTML]{B7B7B7}}                 \\

\multicolumn{1}{c|}{}                       & \multicolumn{1}{c|}{HHAR}        & \multicolumn{1}{c|}{\uline{51.24 ± 0.32}} & \multicolumn{1}{c}{28.34 ± 0.69} & \multicolumn{1}{c}{\textbf{66.64 ± 1.61}} & 48.99 ± 3.10 \\
\multicolumn{1}{c|}{}                       & \multicolumn{1}{c|}{MobiAct}     & \multicolumn{1}{c|}{70.12 ± 2.31} & \multicolumn{1}{c}{41.32 ± 5.59} & \multicolumn{1}{c}{\textbf{84.73 ± 1.16}} & \uline{78.52 ± 0.75} \\
\multicolumn{1}{c|}{}                       & \multicolumn{1}{c|}{MotionSense} & \multicolumn{1}{c|}{87.99 ± 0.45} & \multicolumn{1}{c}{64.82 ± 2.93} & \multicolumn{1}{c}{\textbf{93.93 ± 0.62}} & \uline{90.28 ± 0.56} \\
\multicolumn{1}{c|}{}                       & \multicolumn{1}{c|}{RealWorld}   & \multicolumn{1}{c|}{\uline{84.62 ± 0.29}} & \multicolumn{1}{c}{58.75 ± 2.47} & \multicolumn{1}{c}{\textbf{89.06 ± 0.28}} & 80.61 ± 1.00 \\
\multicolumn{1}{c|}{}                       & \multicolumn{1}{c|}{UCI}         & \multicolumn{1}{c|}{\uline{92.07 ± 0.53}} & \multicolumn{1}{c}{70.05 ± 3.79} & \multicolumn{1}{c}{\textbf{93.40 ± 0.38}} & 87.04 ± 1.23 \\
\multicolumn{1}{c|}{\multirow{-6}{*}{HART}} & \multicolumn{1}{c|}{PAMAP2}       & \multicolumn{1}{c|}{\uline{63.70 ± 0.72}} & \multicolumn{1}{c}{37.95 ± 0.90} & \multicolumn{1}{c}{\textbf{67.58 ± 0.84}} & 57.61 ± 2.66
\\ 
\hline
\end{tabular}
\label{table:frozenAllResults}
\end{table}

Table~\ref{table:frozenAllResults} shows the results with the model's feature extractor frozen on each left-out dataset. We observe that MAE consistently outperforms all the other methods on both architectures on all datasets, showcasing its effectiveness as a self-supervised learning approach. Data2vec and the supervised baseline share similar results. 
SimCLR, with the iSPL architecture, performs competitively, having the second-best results on the PAMAP2, MotionSense, and Mobiact datasets, but drastically falls behind and performs the worst on the other datasets. When used with HART architecture, SimCLR has the worst overall results among the compared techniques. 

Except for SimCLR, the pre-training method on the transformer-based architecture generally leads to improvements in the recognition rate. This is likely due to the attention mechanism, which greatly benefits the MAE and Data2vec masking process.

\subsubsection{Unfrozen Feature Extractor}
\label{subsec:datasetlevelunfrozen}

 Table~\ref{table:unfrozenAllResults} presents the results of the model's feature extractor unfrozen for different datasets using the iSPL and HART architectures.

\begin{table}[h]
\centering
\caption{Average F1-scores (± std) over 5 random runs with the model's feature extractor \textbf{unfrozen} for each dataset as a downstream task. Best results are reported in \textbf{bold}, while second-best are performances are \uline{underlined} }
\begin{tabular}{cccccc}
Archt.    & Datasets & Supervised     & SimCLR     & MAE    & Data2vec     \\ \hline
\multicolumn{1}{c|}{}  & \multicolumn{1}{c|}{HHAR}         & \multicolumn{1}{c|}{\textbf{95.05 ± 0.42}}          & \multicolumn{1}{c}{\uline{95.01 ± 0.41}}          & \multicolumn{1}{c}{92.84 ± 1.23}          & 92.38 ± 0.60          \\
\multicolumn{1}{c|}{}& \multicolumn{1}{c|}{MobiAct}      & \multicolumn{1}{c|}{\uline{88.85 ± 0.52}}          & \multicolumn{1}{c}{\textbf{89.06 ± 0.33}} & \multicolumn{1}{c}{86.88 ± 0.38}          & 88.06 ± 0.35          \\
\multicolumn{1}{c|}{}    & \multicolumn{1}{c|}{MotionSense}  & \multicolumn{1}{c|}{\textbf{98.18 ± 0.22}} & \multicolumn{1}{c}{\uline{97.92 ± 0.29}}          & \multicolumn{1}{c}{95.81 ± 0.67}          & 96.40 ± 0.39          \\
\multicolumn{1}{c|}{}     & \multicolumn{1}{c|}{RealWorld}    & \multicolumn{1}{c|}{\textbf{91.34 ± 0.35}} & \multicolumn{1}{c}{\uline{90.72 ± 0.31}}          & \multicolumn{1}{c}{89.24 ± 0.41}          & 89.13 ± 0.63          \\
\multicolumn{1}{c|}{}  & \multicolumn{1}{c|}{UCI}          & \multicolumn{1}{c|}{\textbf{97.69 ± 0.08}} & \multicolumn{1}{c}{\uline{97.31 ± 0.41}}          & \multicolumn{1}{c}{95.22 ± 0.58}          & 94.95 ± 0.58          \\
\multicolumn{1}{c|}{\multirow{-6}{*}{iSPL}} & \multicolumn{1}{c|}{PAMAP2}       & \multicolumn{1}{c|}{\uline{73.54 ± 1.07}}          & \multicolumn{1}{c}{\textbf{74.86 ± 2.59}} & \multicolumn{1}{c}{68.82 ± 1.76}          & 70.53 ± 1.29          \\
\multicolumn{6}{l}{\cellcolor[HTML]{B7B7B7}}  \\
\multicolumn{1}{c|}{}         & \multicolumn{1}{c|}{HHAR}         & \multicolumn{1}{c|}{\uline{94.14 ± 0.38}}          & \multicolumn{1}{c}{93.20 ± 0.65}           & \multicolumn{1}{c}{\textbf{94.38 ± 0.46}}          & 94.11 ± 0.59 \\
\multicolumn{1}{c|}{} & \multicolumn{1}{c|}{MobiAct}      & \multicolumn{1}{c|}{\textbf{87.70 ± 0.26}} & \multicolumn{1}{c}{\uline{87.17 ± 0.96}}          & \multicolumn{1}{c}{87.08 ± 0.45}          & 86.70 ± 0.47          \\
\multicolumn{1}{c|}{} & \multicolumn{1}{c|}{MotionSense}  & \multicolumn{1}{c|}{\textbf{97.84 ± 0.14}} & \multicolumn{1}{c}{\uline{97.55 ± 0.20}}           & \multicolumn{1}{c}{97.11 ± 0.28}          & 97.20 ± 0.22          \\
\multicolumn{1}{c|}{} & \multicolumn{1}{c|}{RealWorld}    & \multicolumn{1}{c|}{\textbf{91.58 ± 0.12}}          & \multicolumn{1}{c}{90.92 ± 0.25}          & \multicolumn{1}{c}{\uline{91.37 ± 0.08}} & 91.00 ± 0.28          \\
\multicolumn{1}{c|}{}& \multicolumn{1}{c|}{UCI}          & \multicolumn{1}{c|}{\textbf{96.78 ± 0.17}}          & \multicolumn{1}{c}{96.17 ± 0.36}          & \multicolumn{1}{c}{\uline{96.32 ± 0.68}} & 95.30 ± 0.95          \\
\multicolumn{1}{c|}{\multirow{-6}{*}{HART}} & \multicolumn{1}{c|}{PAMAP2}       & \multicolumn{1}{c|}{68.98 ± 0.65}          & \multicolumn{1}{c}{70.99 ± 0.93}   & \multicolumn{1}{c}{\uline{71.14 ± 1.13}}          & \textbf{71.54 ± 1.00} \\
\hline \end{tabular}
\label{table:unfrozenAllResults}
\end{table}

In the unfrozen scenarios, there is no clear winner between the pre-training methods. Differences in performances are generally below a 1\% margin between the best and performing techniques on both architectures. Given enough data for the fine-tuning, such as simulated here in the data-rich fine-tuning environment, all model's weight can converge to very similar performances when used with the same architecture.

\subsection{Results considering labeled data scarcity scenarios}

Previous results focused on the ideal case where a high amount of data is available for fine-tuning in the target domain, and it is possible to perform training for many epochs. Nevertheless, this is not likely the case when we downstream the training onto a single user's mobile/wearable device, which is likely resource-constrained. In this section, we present our findings when labeled data is scarce (i.e., 1, 2, 5, 10, 20, 50, 100 samples of each class available for training) and the training process is constrained (from 200 epochs to only 50 epochs). We report our findings only on the RealWorld and MotionSense datasets for presentation purposes. However, the complete results on all seven datasets are also given in the shared code repository.

\subsubsection{Frozen Feature Extractor}

\begin{figure}[h]
    \centering
    \begin{subfigure}{.50\textwidth}
        \centering
        \includegraphics[width=\linewidth]{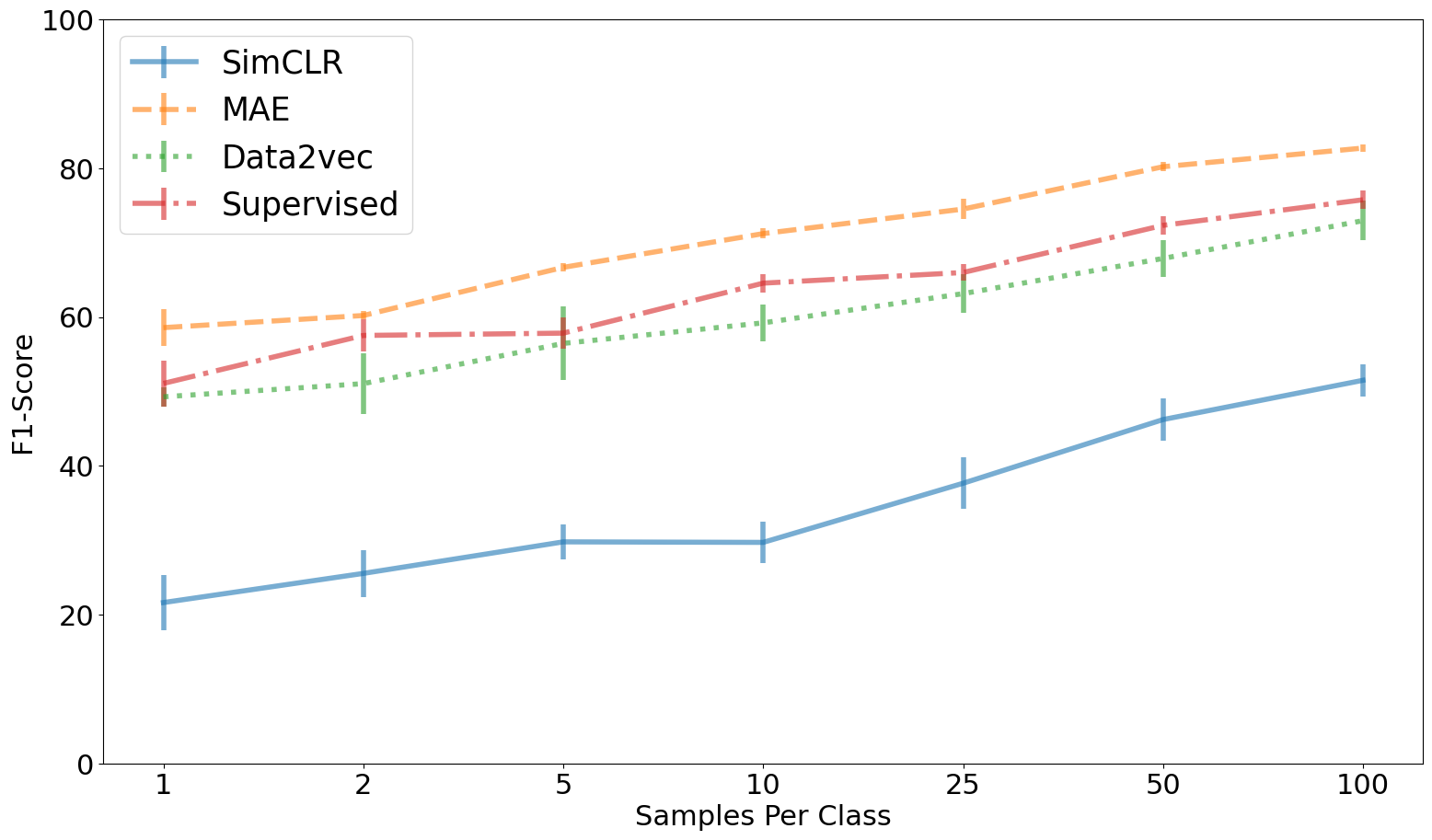}
        \caption{HART - RealWorld}
        \label{fig:frozenFewHARTRealworld}
    \end{subfigure}%
    \begin{subfigure}{.50\textwidth}
        \centering
        \includegraphics[width=\linewidth]{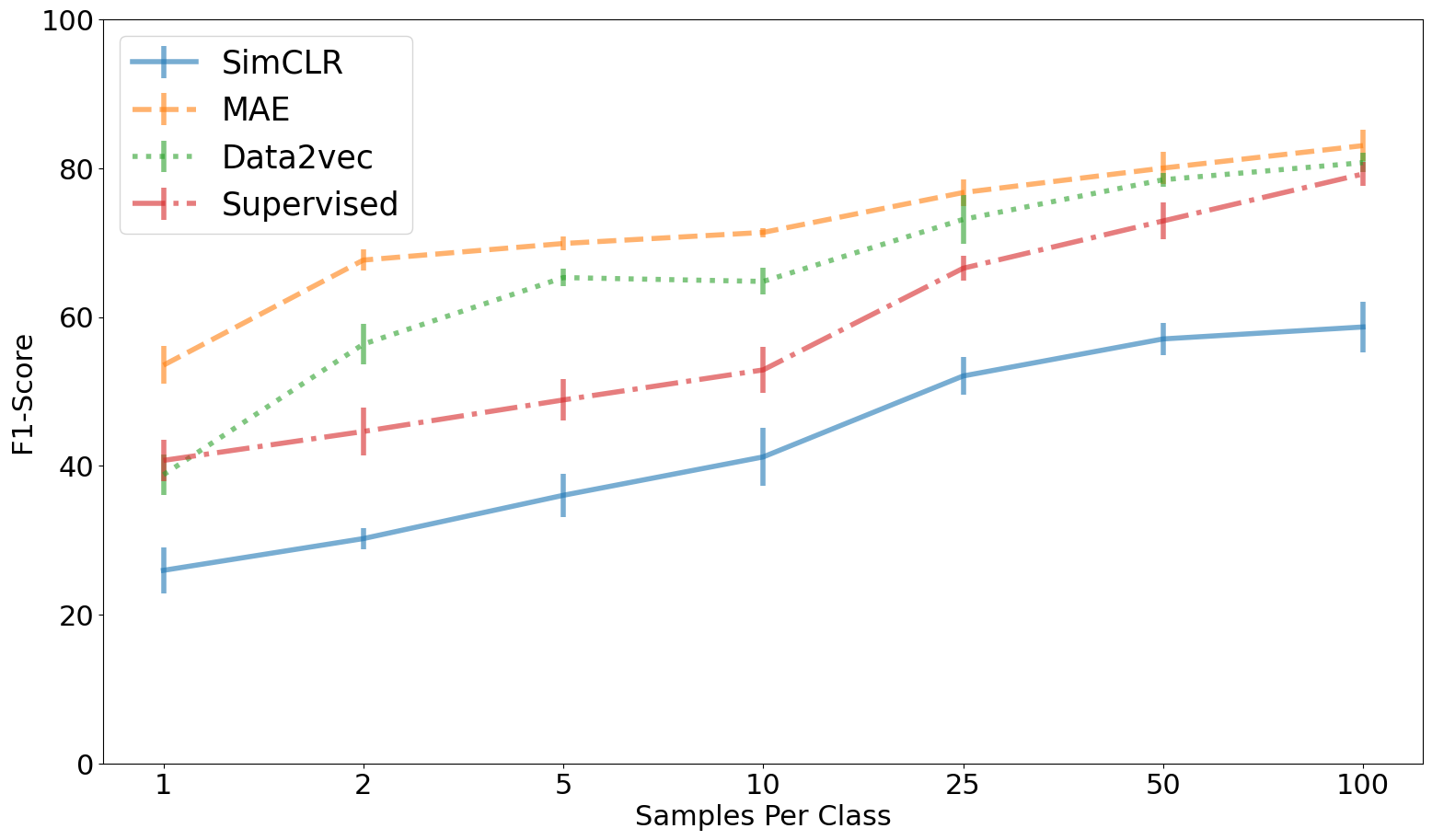}
        \caption{HART - MotionSense}
        \label{fig:frozenFewHARTNMotionSense}
    \end{subfigure}

    \begin{subfigure}{.50\textwidth}
        \centering
        \includegraphics[width=\linewidth]{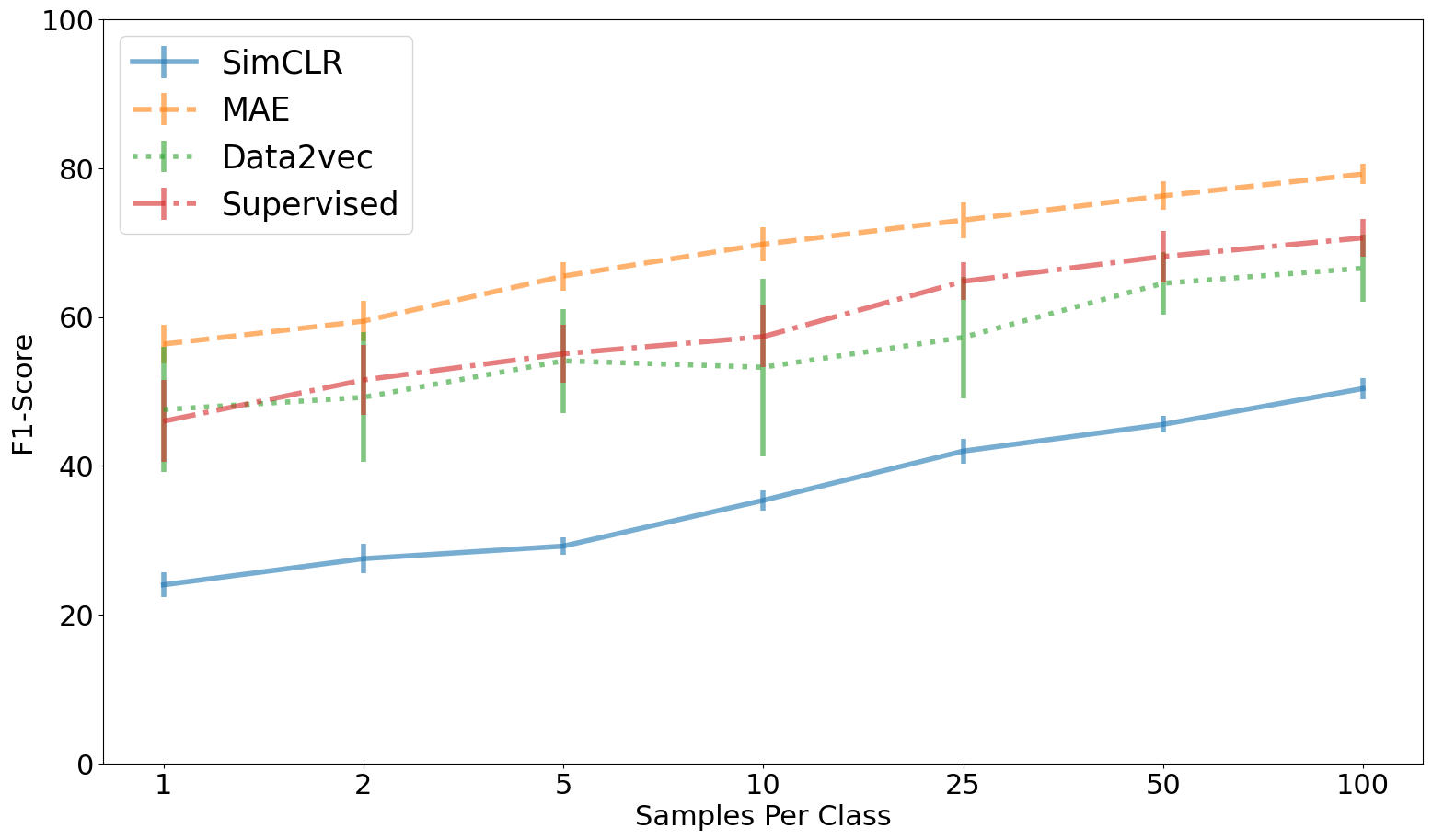}
        \caption{iSPL - RealWorld}
        \label{fig:frozenFewISPLRealworld}
    \end{subfigure}%
    \begin{subfigure}{.50\textwidth}
        \centering
        \includegraphics[width=\linewidth]{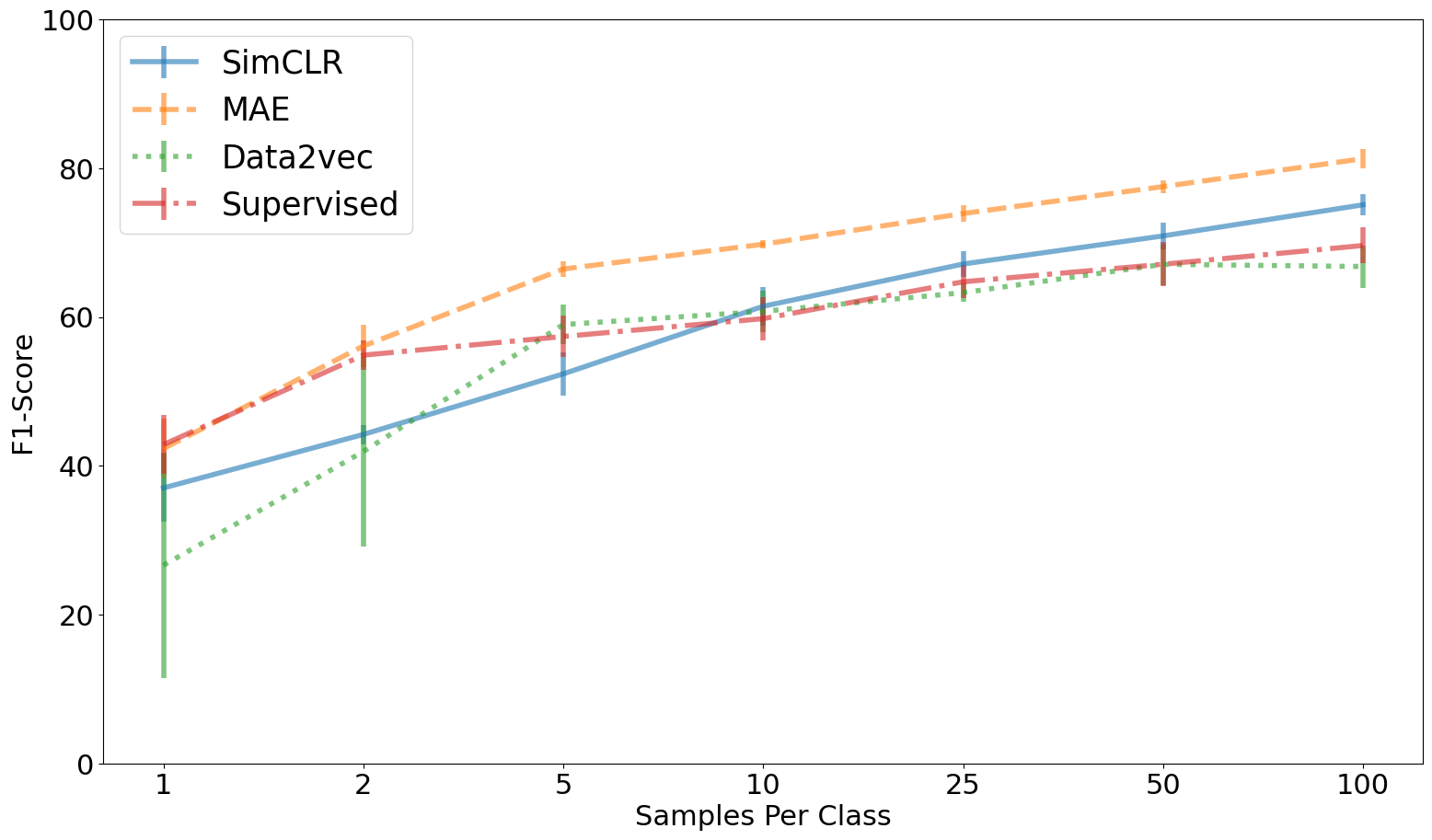}
        \caption{iSPL - MotionSense}
        \label{fig:frozenFewISPLMotionSense}
    \end{subfigure}
    \caption{Comparison with the feature extractor frozen  }
    \label{fig:frozenFew}
\end{figure}

Figure~\ref{fig:frozenFew} shows the performance of each SSL method with a frozen feature extractor during fine-tuning and a small number of samples per activity class. 

MAE has performed the best overall on both datasets and architectures, obtaining performances well above 60\% in the F-score with only 10 samples per class in the considered scenarios. On the contrary, SimCLR performed very poorly, especially on the RealWorld datasets on both architectures where the method's performance is far below other techniques. Data2vec and the Supervised approach performed very comparatively, often interchanging second and third place in performance.

\subsubsection{Unfrozen Feature Extractor}
\begin{figure}[h]
    \centering
    \begin{subfigure}{.50\textwidth}
        \centering
        \includegraphics[width=\linewidth]{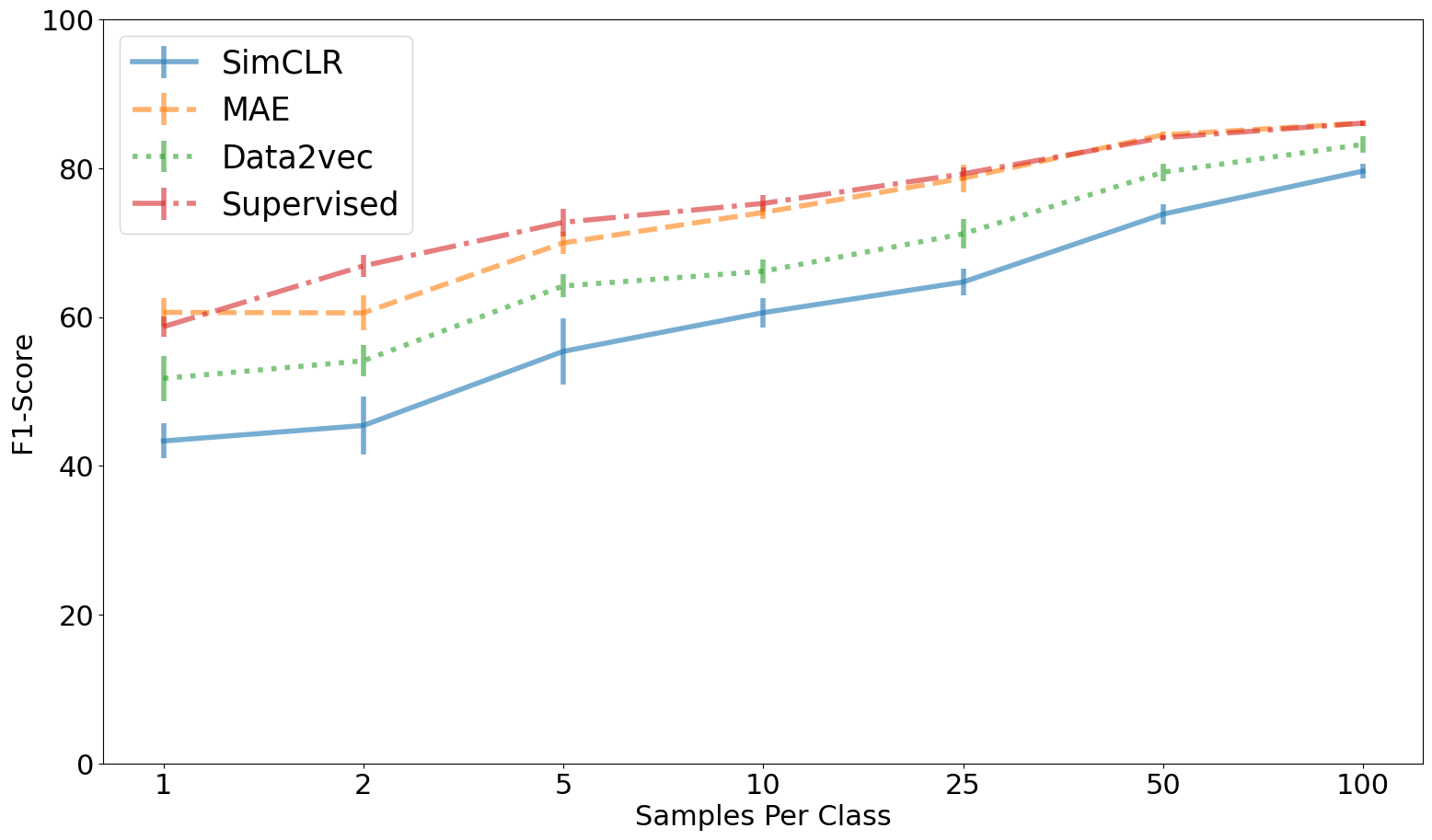}
        \caption{HART - RealWorld}
        \label{fig:unfrozenFewHARTRealworld}
    \end{subfigure}%
    \begin{subfigure}{.50\textwidth}
        \centering
        \includegraphics[width=\linewidth]{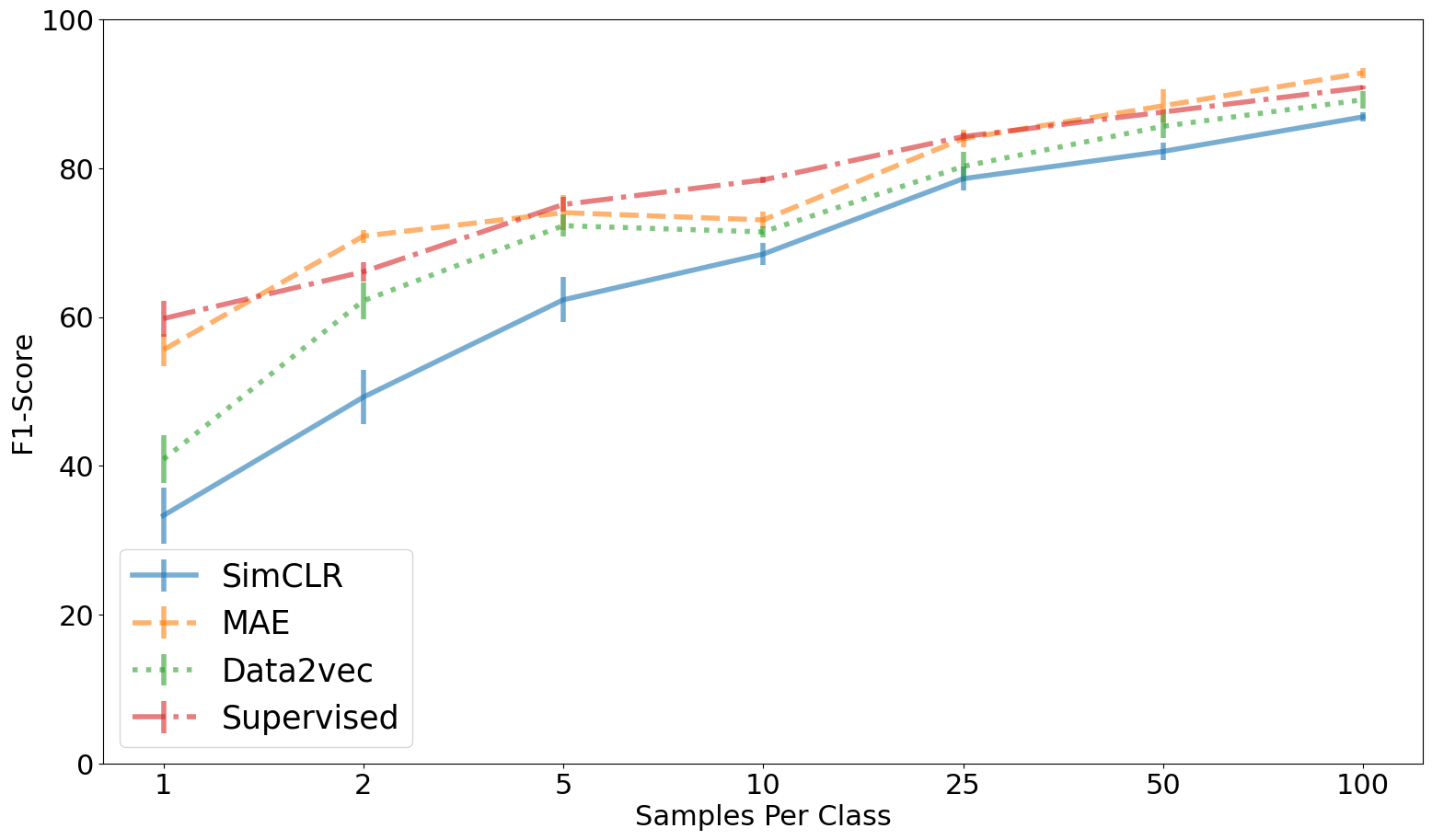}
        \caption{HART - MotionSense}
        \label{fig:unfrozenFewHARTNMotionSense}
    \end{subfigure}

    \begin{subfigure}{.50\textwidth}
        \centering
        \includegraphics[width=\linewidth]{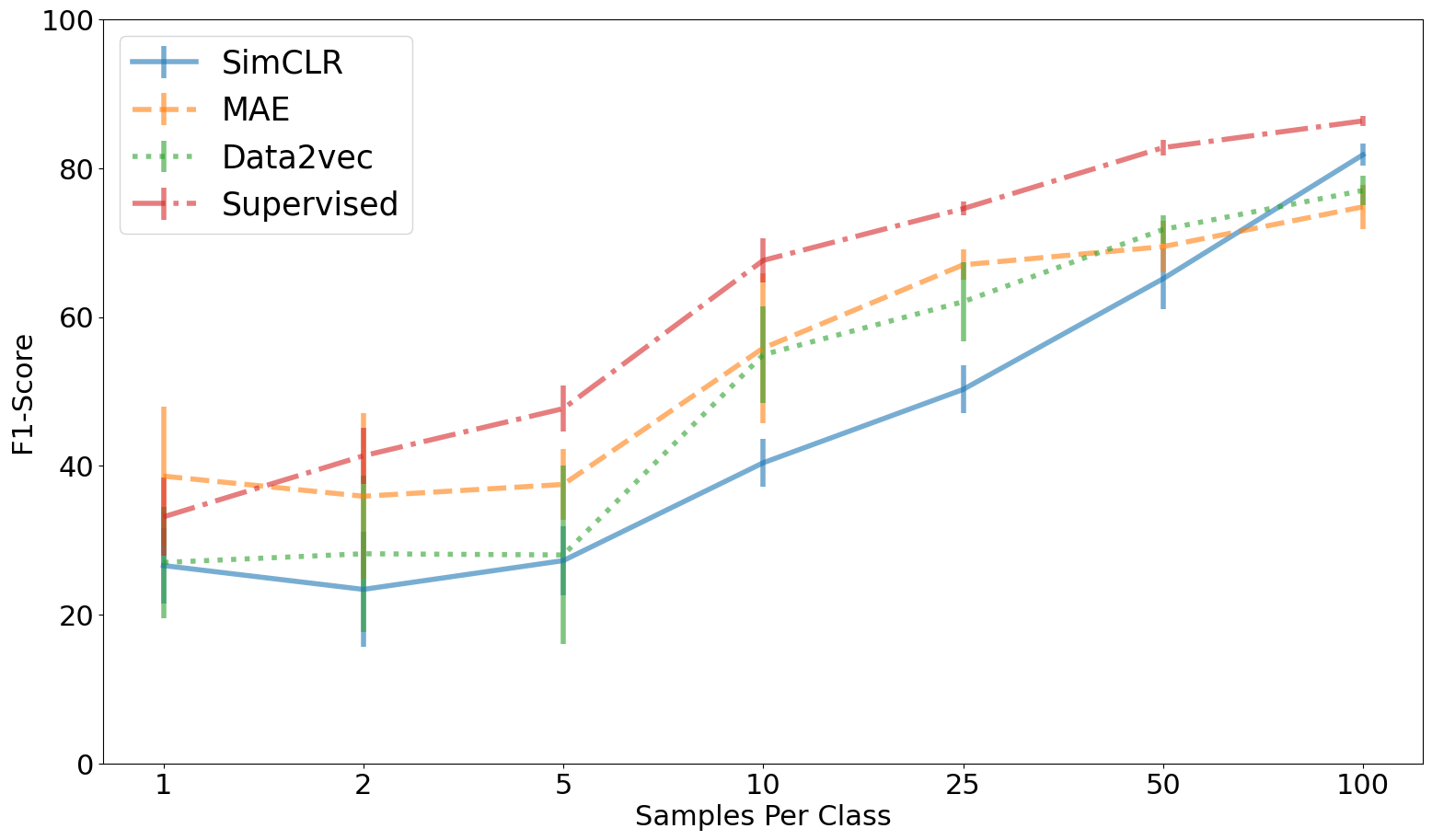}
        \caption{iSPL - RealWorld}
        \label{fig:unfrozenFewISPLRealworld}
    \end{subfigure}%
    \begin{subfigure}{.50\textwidth}
        \centering
        \includegraphics[width=\linewidth]{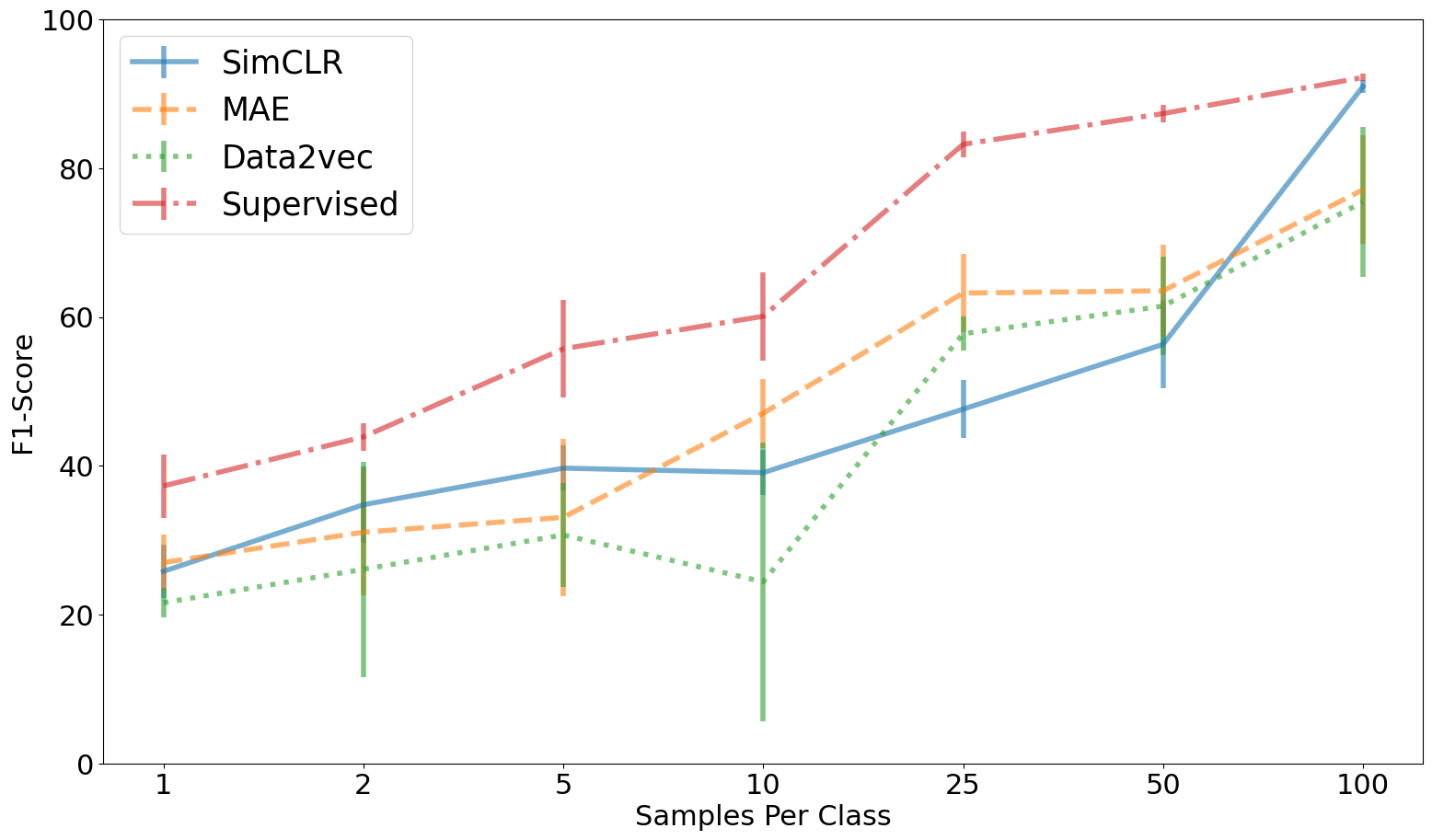}
        \caption{iSPL - MotionSense}
        \label{fig:unfrozenFewISPLMotionSense}
    \end{subfigure}
    \caption{Comparison with the feature extractor unfrozen  }
    \label{fig:unfrozenFew}
\end{figure}

Figure~\ref{fig:unfrozenFew} compares the SSL techniques when the feature extractor is unfrozen. While we observe that the best results are generally obtained by pre-training the classifier in a supervised fashion (that is an unrealistic scenario), MAE was ranked 2nd in all cases except with results not far behind.

The worst-performing results, in general, were models that were pre-trained with SimCLR. Data2vec often performed in the middle ground compared to other techniques, except when we used the iSPL model on the MotionSense dataset, where data2vec performed poorly between the 1 - 10 samples per class range.

\subsection{Efficiency}
\label{subsec:efficiency}

Table~\ref{tab:parameters_flops} presents the number of FLOPs and parameters for the two architectures after the needed adaption for each pre-training technique. Firstly, we show that SimCLR required a more significant number of FLOPs as the encoder in the learning pipeline was used multiple times for each transformation. Also, we found that MAE and data2vec significantly required fewer FLOPs, again with the iSPL model. HART required much fewer FLOPs than iSPL for almost all methods. In terms of parameter counts, we see an opposite trend where the parameters of the supervised and SimCLR are about half less than those of other methods. Due to the encoder-decoder and siamese configuration of MAE and Data2vec, their number of parameters are about double that of the supervised and SimCLR.



\begin{table}[]
\centering
\caption{Efficiency comparison of the methods and architectures FLOPs and parameters}
\begin{tabular}{ccrrrr}
                                                  & Archt.                    & \multicolumn{1}{c}{Supervised}   & \multicolumn{1}{c}{SimCLR}       & \multicolumn{1}{c}{MAE}         & \multicolumn{1}{c}{Data2vec} \\ \hline
\multicolumn{1}{c|}{}                             & \multicolumn{1}{c|}{iSPL} & \multicolumn{1}{r|}{338,528,096} & \multicolumn{1}{r|}{677,467,252} & \multicolumn{1}{r|}{72,119,324} & 45,566,144                \\
\multicolumn{1}{c|}{\multirow{-2}{*}{FLOPs}}      & \multicolumn{1}{c|}{HART} & \multicolumn{1}{r|}{30,480,616}  & \multicolumn{1}{r|}{15,001,422}  & \multicolumn{1}{r|}{24,946,174} & 29,482,584                  \\
\multicolumn{6}{l}{\cellcolor[HTML]{B7B7B7}}                                                                                                                                                                    \\
\multicolumn{1}{c|}{}                             & \multicolumn{1}{c|}{iSPL} & \multicolumn{1}{r|}{1,327,714}   & \multicolumn{1}{r|}{1,430,282}   & \multicolumn{1}{r|}{2,178,144}  & 2,855,776                    \\
\multicolumn{1}{c|}{\multirow{-2}{*}{Parameters}} & \multicolumn{1}{c|}{HART} & \multicolumn{1}{r|}{1,450,018}   & \multicolumn{1}{r|}{1,330,890}   & \multicolumn{1}{r|}{3,280,440}  & 2,473,040                   
\end{tabular}
\label{tab:parameters_flops}

\end{table}

\begin{figure}[h]
    \centering
    \begin{subfigure}{.50\textwidth}
        \centering
        \includegraphics[width=\linewidth]{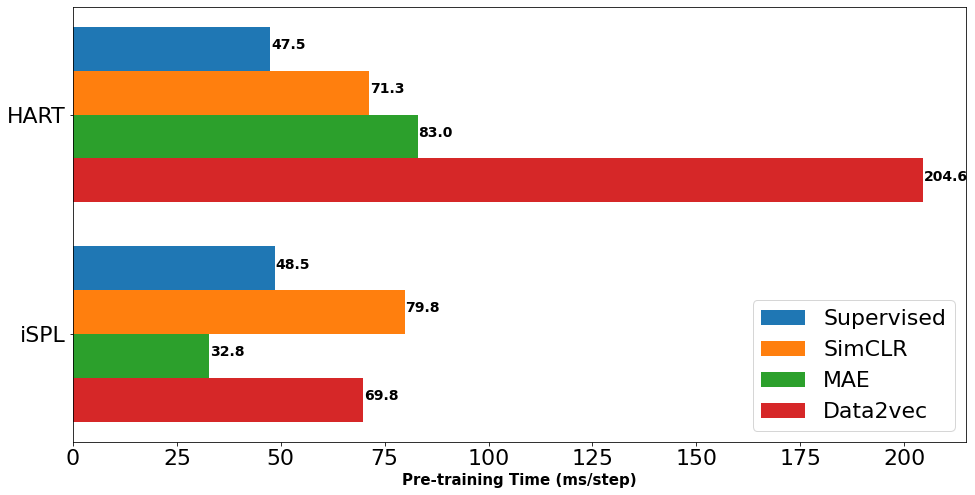}
        \caption{Pre-train Time}
        \label{fig:pretrain_efficiency}
    \end{subfigure}%
    \begin{subfigure}{.50\textwidth}
        \centering
        \includegraphics[width=\linewidth]{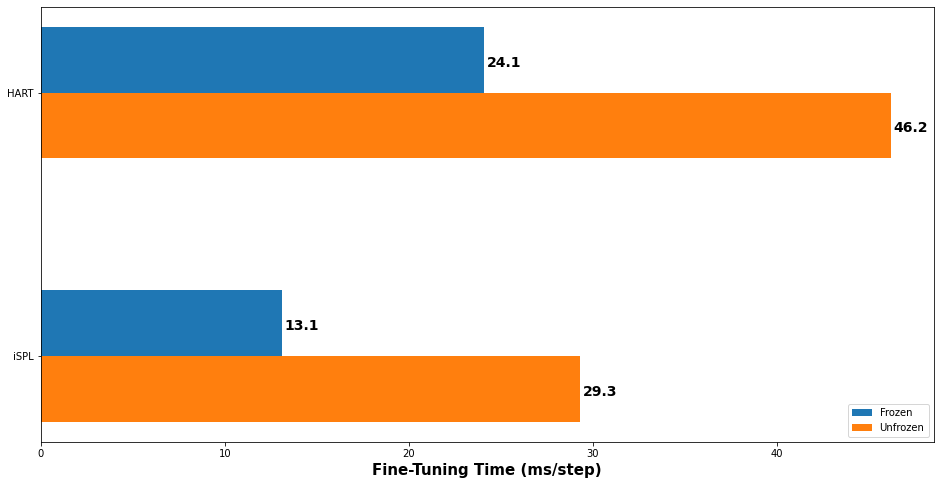}
        \caption{Fine-tuning Time}
        \label{fig:downstream_efficiency}
    \end{subfigure}
    
    \caption{Measurements of training time per step (ms/step) for each combination of techniques and architectures }
\end{figure}

Afterward, we measure and compare the training efficiency of the different techniques for a single step/batch and report the average in milliseconds per step in Figure~\ref{fig:pretrain_efficiency}. Data2vec with HART required 204.6 ms/step on average, the longest of all other evaluations. This result is caused by the manual update of the teacher's weight for each layer after every batch through the moving average in our implementation (we note that HART has more players than ISPL). MAE with ISPL, relatively, had the shortest training time, around 32.8 ms/step, even faster than the supervised training, but was slower when used with HART. Lastly, we found that SimCLR had a more consistent and moderate training time between the two architectures (71.3 ms/step on HART and 79.8 ms/step on ISPL).



When we look at the training efficiency during down-streaming, we compared the two architectures' training time when the feature extractor is frozen and unfrozen. The frozen scenario is expected to be significantly faster as we will not need to back-propagate the entire model during the training stage. This property is true, as shown in figure \ref{fig:downstream_efficiency}, where when the feature extractor is frozen, we usually see about a $2$ times speed-up over the unfrozen case in average training time per batch.

Overall, while SSL methods do introduce overhead compared to the supervised learning approach, they have the advantage of preceding the need for labeled data for pre-training. 


\section{Discussion}
\label{sec:discussion}
The gains from each of the three SSL techniques are best measured when the weights from the pre-training are unchanged during fine-tuning such that the feature extractor is frozen. In this aspect, the feature extractor trained through MAE had clear superiority against other techniques. Even in terms of training efficiency, compared to other SSL methods, MAE was the second fastest on HART and the fastest, even against the supervised approach, on iSPL. 

Afterward, we saw that data2vec performed competitively but could not match the MAE performance results in our study. Data2vec had the longest training time of all when used with HART. 

Lastly, we found that while SimCLR performed well when paired with a CNN architecture, it is the least effective technique amongst the studied SSL methods when paired with a transformer-based architecture. The results align with recent findings where other domains have shifted to generative and predictive approaches~\cite{he2022masked,baevski2022data2vec}.

In the unfrozen setting, most performance was equivalent across the SSL methods compared. In the data-scarce environment, we saw that self-supervised pre-training allowed the model to perform better in extremely low sample ranges when compared to starting from random, and results converged to equivalent performance with more training data.

Training with the feature extractor frozen offers gains in both efficiency and performance in extreme data-scarce scenarios. When there are fewer than 25 samples per class, training the feature extractor frozen showed a significant gap in performance over the unfrozen scenario. Lastly, the frozen case allowed faster training time, more than double the gain compared to the unfrozen setting, with the addition of the natural computation gains due to not having to back-propagate the entire model during the learning. 

Finally, our findings have shown a strong correlation between specific architectures and the SSL method. In particular, SimCLR performed relatively well when used with CNN-based iSPL. When SimCLR was adopted with HART, there was a significant drop in the F-score, a 25\% drop in several cases. On the other hand, MAE and data2vec performed more when used with transformer-based HART as each frame became contextualized embeddings of other frames.

\section{Conclusion}
\label{sec:conclusion}

In this paper, we presented a comprehensive evaluation of three state-of-the-art Self-Supervised Learning (SSL) techniques for the problem of wearable Human Activity Recognition (HAR). In order to take into account the recent evolution of SSL techniques from other domains and compare them to SimCLR with a reproducible benchmark, we adapted and evaluated three techniques from three families of SSL, namely Contrastive Learning (SimCLR), Generative Learning (MAE), and Predictive Learning (data2vec). To the best of our knowledge, this is the first time that MAE and Date2Vec are assessed in the HAR domain. An extensive evaluation involving several datasets and two main classification models (CNN-based and transformer-based) showed that MAE emerges as the most robust technique for the domain in the frozen case, surpassing all other approaches, including supervised pre-training methods. Moreover, in the case of full fine-tuning (unfrozen case), MAE stays in part with classical supervised techniques with transformer models. Finally, in the case of low resource settings (e.g., low amount of labeled data), the MAE technique also shows an excellent transfer in extreme cases. Finally, the code and pre-trained models have been made publicly available to encourage reproducibility.

Although the LODO technique has been a way to mitigate the lack of available data, the current study did not involve a massive amount of data as in other computing domains (e.g., vision or text). We hope to reach such a comparable amount as emerged with the release of the recent Capture-24 ($\sim$ 2,500 hr data)~\cite{chan2021capture} and BioBank ($\sim$ 100,000 participants)\cite{doherty2017large} datasets, even if they include a reduced variety of sensors. Moreover, availability issues may still prevent free access to such data.
Another area for improvement is that our work has been conducted on offline experiments, and the behavior of such models in real settings needs to be evaluated. We plan to investigate how these techniques can be applied with active learning~\cite{arrotta2023selfact}. Finally, the study was restricted to three SSL techniques and two models. We plan to extend the coverage by considering more SSL techniques such as JEPA~\cite{assran2023self}.

\section{Acknowledgement}
\label{sec:acknowledgement}
This work has been partially funded by Naval Group, by MIAI@Grenoble Alpes (ANR-19-P3IA-0003), and granted access to the HPC resources of IDRIS under the allocation 2023-AD011013233R1 made by GENCI.

Part of this research was also supported by projects SERICS (PE00000014) and by project MUSA – Multilayered Urban Sustainability Action,  funded by the European Union – NextGenerationEU, under the National Recovery and Resilience Plan (NRRP) Mission 4 Component 2 Investment Line 1.5: Strengthening of research structures and creation of R\&D “innovation ecosystems”, set up of “territorial leaders in R\&D”.

\subsection{Data availability}

All the datasets used during the study are described in Section~\ref{sec:eval_dataset} and are available through the following links : 
\begin{itemize}
    \item UCI: \url{https://archive.ics.uci.edu/ml/datasets/human+activity+recognition+using+smartphones}
    \item Motionsense: \url{https://github.com/mmalekzadeh/motion-sense/tree/master/data}
    \item HHAR: \url{http://archive.ics.uci.edu/ml/datasets/Heterogeneity+Activity+Recognition}
    \item RealWorld (RW): \url{https://sensor.informatik.uni-mannheim.de/#dataset_dailylog}
    \item MobiAct: \url{https://bmi.hmu.gr/the-mobifall-and-mobiact-datasets-2/}
    \item PAMAP2: \url{https://archive.ics.uci.edu/dataset/231/pamap2+physical+activity+monitoring}
\end{itemize}

\bibliographystyle{unsrt}  
\bibliography{bibfile}

\end{document}